\documentclass[aps,prb,twocolumn,superscriptaddress]{revtex4}

\usepackage[dvips]{graphicx}
\usepackage[dvips]{color}
\usepackage{amsmath}

\renewcommand{\Vec}[1]{\mbox{\boldmath$#1$}}

\begin{document}

\title{Faraday rotation in bilayer and trilayer graphene in the quantum Hall regime}

\author{Takahiro Morimoto}
\affiliation{Condensed Matter Theory Laboratory, Riken, Saitama, 351-0198, Japan}
\author{Mikito Koshino}
\affiliation{Department of Physics, Tohoku University, Sendai, 980-8578, Japan}
\author{Hideo Aoki}
\affiliation{Department of Physics, University of Tokyo, Hongo, 
Tokyo 113-0033, Japan}

\date{\today}

\begin{abstract}
Optical Hall conductivity, as directly related to Faraday rotation, 
is  theoretically studied for bilayer and trilayer graphene.  
In bilayer graphene, the trigonal warping of the band dispersion
greatly affects the resonance structures in Faraday rotation 
not only in the low-energy region where small Dirac cones emerge, 
but also in the higher-energy parabolic bands 
as a sequence of satellite resonances.   
In ABA-stacked trilayer, the resonance spectrum is 
a superposition of effective monolayer and bilayer contributions with band gaps, 
while ABC trilayer exhibits a distinct spectrum 
peculiar to the cubic-dispersed bands with a strong trigonal warping,
where the signals associated with low-energy Dirac cones 
should be directly observable
owing to a large Lifshitz transition energy ($\sim 10 \mbox{ meV}$).
\end{abstract}


\maketitle


\section{Introduction}
The observation of the anomalous quantization of $\sigma_{xy} = \frac{4e^2}{h} (n+ \frac 1 2)$ in graphene quantum Hall system has established the existence of massless Dirac-quasiparticle in graphene\cite{Nov05,Zha05},
and kicked off an increasing fascination with the massless Dirac physics of 
graphene.  Optical properties have also been studied, where a point of 
interest is 
an unusual optical  selection rule.   Specifically, 
the optical Hall conductivity 
$\sigma_{xy}(\omega)$ 
is an interesting quantity to look at,
since it is an ac-extension of the Hall conductivity 
which has a topological nature.

The optical Hall conductivity has been theoretically studied for graphene \cite{gusynin-sxy,morimoto-opthall,Fialkovsky09}, 
graphite \cite{Falkovsky-graphite11}
and topological insulators\cite{tse-macdonald10}.  
Experimentally, the optical Hall conductivity is measurable through Faraday rotation, 
since the Faraday rotation angle $\Theta_H$ is proportional to  $\sigma_{xy}(\omega)$ in quantum Hall regime [where 
$(n_0+n_s) \gg |\sigma_{xx}(\omega) \pm \sigma_{xy}(\omega)|/(c\epsilon_0)$] as
\begin{equation}
\Theta_H  \simeq
\frac{1}{(n_0+n_s)c\varepsilon_0} \mbox{Re}\,\sigma_{xy}(\omega),
\label{faraday}
 \end{equation}
where $n_0 (n_s)$ is the refractive index of the air (substrate)\cite{oconnell82}, $c$ the velocity of light, and $\varepsilon_0$ the dielectric constant of 
vacuum.
Faraday rotation experimentally starts to be measured both for 2DEG \cite{ikebe-THz10}
and for graphene \cite{crassee2010giant} in the quantum Hall regime.
Due to  the Dirac nature of quasiparticles in graphene,
the optical Hall conductivity in graphene reflects
the Dirac Landau level (LL) structure 
($\varepsilon_n={\rm sgn}(n)\sqrt{n} \hbar \omega_c$)
and an unusual selection rule ($ |n| \leftrightarrow |n|+1$),
with Landau index $n$.

\begin{figure}[tb]
\begin{center}
\includegraphics[width=1.0\linewidth]{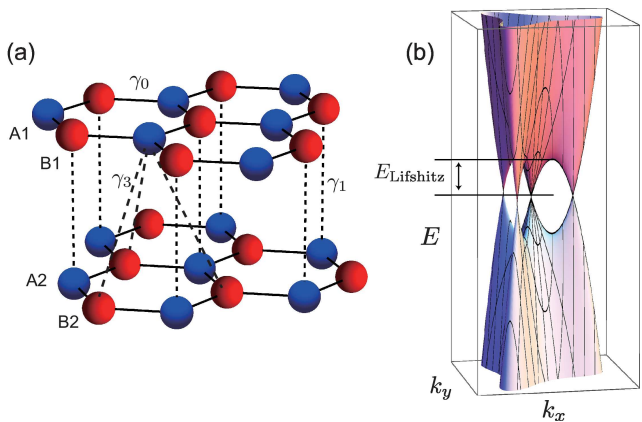}
\end{center}
\caption{
(a) Lattice structure of bilayer graphene with 
A/B sublattice carbons indicated in blue/red.
The unit cell consists of $A_1$ and $B_1$ carbon atoms in the top layer and  $A_2$ and $B_2$ carbon atoms in the bottom.  
$\gamma_0$: the nearest-neighbor ($A_1 \leftrightarrow B_1, A_2 \leftrightarrow B_2$) hopping within each layer, 
$\gamma_1$: the vertical ($B_1 \leftrightarrow A_2 $) interlayer hopping, 
and $\gamma_3$: an oblique ($A_1 \leftrightarrow B_2 $) 
interlayer hopping.  Two layers are separated by $d=0.334$ nm.
(b) Low-energy band dispersion of bilayer graphene with the trigonal warping effect with a Lifshitz transition where the topology of Fermi surface changes from 
four Dirac cones to one, trigonally warped parabolic dispersion.
}
\label{bilayer-lattice}
\end{figure}

Studies so far have focused on the optical Hall conductivity
in monolayer graphene quantum Hall system.  In the 
physics of graphene, 
there are growing interests in bilayer and trilayer graphene systems, 
since their 
electronic structures are distinct from that of monolayer graphene.
For bilayer graphene the interlayer coupling between the two graphene sheets ($\gamma_1$ in Fig.\ref{bilayer-lattice}) changes 
the monolayer's Dirac cone into 
two parabolic bands touching at the K+ and K- points 
in the Brillouin zone. \cite{nov-bilayer,mccann-falko}
To be more precise, if we take account of 
the second-neighbor interlayer hopping ($\gamma_3$ in Fig.\ref{bilayer-lattice}) a trigonal warping is induced on the band dispersion (Fig.\ref{bilayer-lattice}(b)).  
There, the parabolic bands touching at the Dirac points are reformed into four Dirac cones in the low-energy region.  
These are connected to the parabolic bands at a higher energy, 
at which a Lifshitz transition in the topology of Fermi surface takes place.  

It is then an interesting question to ask how these will 
affect optical responses.  
While the optical {\it longitudinal} conductivity $\sigma_{xx}(\omega)$ is discussed in \cite{abergel-falko,carbotte-sxx-bilayer}, and experimentally utilized for determining the band structure \cite{determining-bilayer-band,band-asymmetry-bilayer}, we propose here that 
the optical Hall conductivity $\sigma_{xy}(\omega)$ should be 
a good probe for multi-layer graphene in the quantum Hall regime.
Trilayer graphene, the next in the series of multilayer graphene, 
harbors another interest, since it 
comes with two different stacking orders, ABA and ABC.
The low-energy dispersion of ABA trilayer consists of 
a monolayer-like Dirac cone and 
a bilayer-like parabolic bands,
which are gapped due to the absence of inversion symmetry 
in the lattice structure.  
\cite{guinea-stacks06,latil-2006,partoens2006graphene,lu2006influence,aoki-amawashi-2007,koshino-ando07,koshino-ando08,koshino-aba-2009,koshino2009electronic,koshino-LL-2011,sena-trilayer,jarillo-trilayer}
On the other hand, ABC trilayer has a pair of cubic-dispersed bands
with a larger trigonal warping effect than in bilayer graphene,
\cite{guinea-stacks06,latil-2006,lu-2006-abc,aoki-amawashi-2007,koshino-abc-2009,fang-abc10} 
for which LL splitting due to Lifshitz transition is reported \cite{bao-abc2011}.
Thus the interest for trilayer is how the stacking 
dependence appears in the  optical Hall conductivity.

These have motivated us to study in this paper the 
optical (Hall as well as longitudinal) 
conductivities in bilayer and trilayer graphene systems.  
We shall show for bilayer graphene that 
the Lifshitz transition accompanied by the trigonal warping 
greatly affects the resonance structures in Faraday rotation 
not only on low-energy scale where Dirac cones emerges 
but also in the higher-energy range with parabolic bands as a sequence of satellite resonances,
whose weight can become large in the vicinity of the Lifshitz transition, hence should be experimentally measurable.   
Thus, while it would be difficult to directly observe the Lifshitz transition and low-energy Dirac cones due to the tiny energy scale of  the trigonal 
warping ($\sim 1 \mbox{ meV}$) ,
the optical conductivities should provide an experimentally accessible 
way to detect the trigonal warping as additional resonance structure.

For trilayer graphene, on the other hand, we shall show that the optical conductivities 
are significantly affected by the difference in the stacking order.
In ABA trilayer, the resonance spectrum is 
a superposition of effective monolayer and bilayer contributions with band gap,
while ABC trilayer exhibits a distinct spectrum 
peculiar to the cubic-dispersed bands.
In the latter, the trigonal warping effect is strong with 
a larger Lifshitz transition energy ($\sim 10 \mbox{ meV}$),
so that we predict here 
the signals associated with low-energy Dirac cones 
should be directly observable.

\begin{figure*}[tb]
\begin{center}
\includegraphics[width=0.8\linewidth]{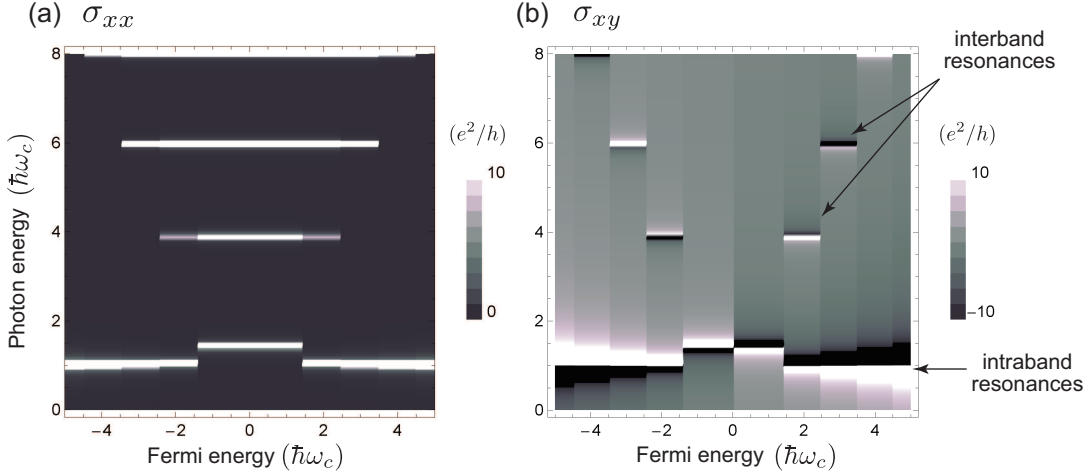}
\end{center}
\caption{For bilayer graphene QHE system without the trigonal warping effect ($v_3=0$), we show 
(a) the optical longitudinal conductivity $\sigma_{xx}(\epsilon_F,\omega)$, and 
(b) optical Hall conductivity $\sigma_{xy}(\epsilon_F,\omega)$, 
grey-scale plotted 
against the Fermi energy $\epsilon_F$ and the frequency $\omega$.
}
\label{wo-warping}
\end{figure*}

\begin{figure*}[tb]
\begin{center}
\includegraphics[width=0.95\linewidth]{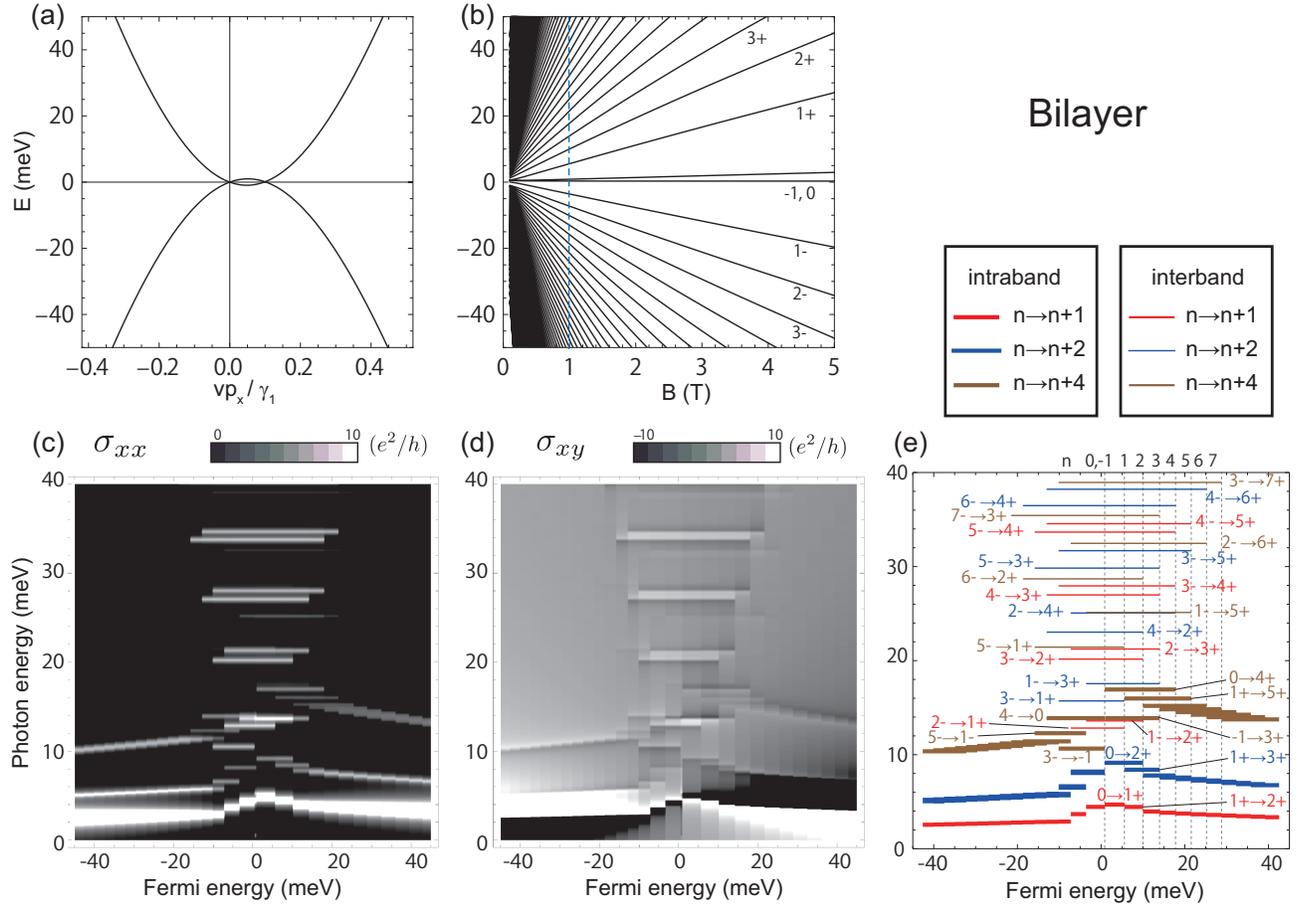}\\
\end{center}
\caption{ 
(a) Low-energy band structure, and (b) Landau levels against magnetic field for bilayer graphene with the trigonal warping included.  
(c) The longitudinal $\sigma_{xx}(\epsilon_F,\omega)$ and (d) Hall $\sigma_{xy}(\epsilon_F,\omega)$ 
gre-scale plotted against the Fermi energy $\epsilon_F$ and the frequency $\omega$
for a magnetic field $B=1T$ (a vertical dashed line in (b)).  
(e) A diagram indicating allowed resonances in $\sigma_{xy}$.  
Vertical dotted lines indicate the Landau levels.
Note that "-1" (n=-1) and "1-" (n=1, s=-1)" are different.
}
\label{Fig-bi2}
\end{figure*}

\begin{figure}[tb]
\begin{center}
\includegraphics[width=0.95\linewidth]{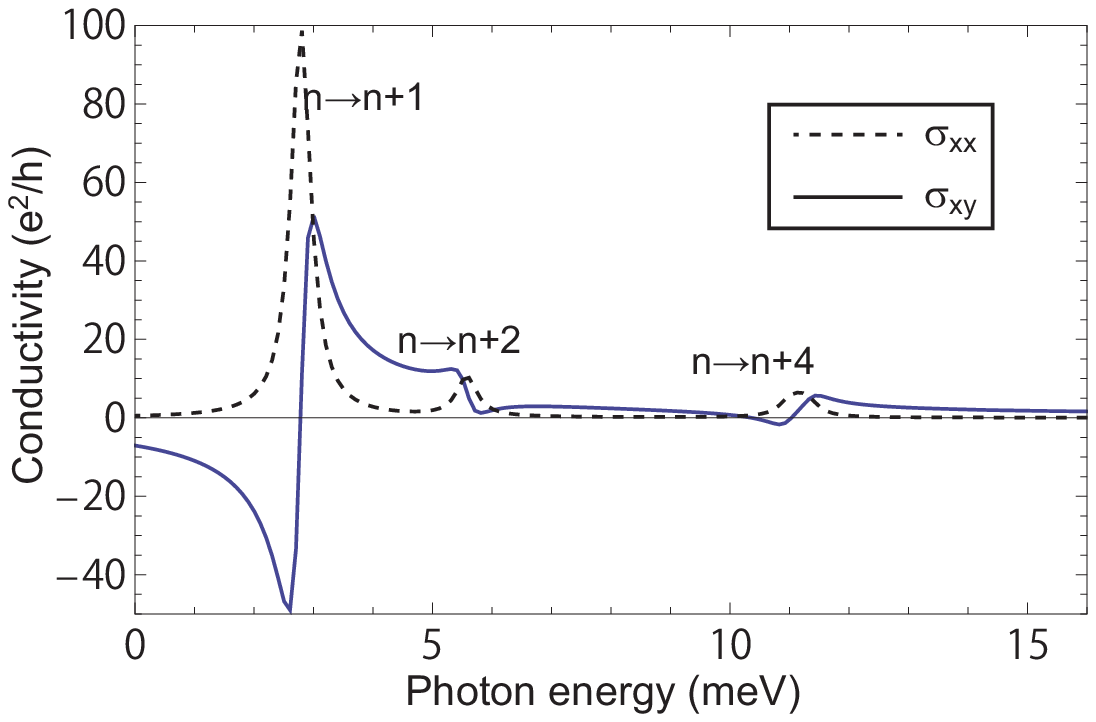}\\
\end{center}
\caption{ 
Optical longitudinal ($\sigma_{xx}$; dashed line) and Hall ($\sigma_{xy}$; solid line) conductivities plotted against the frequency $\omega$
for a magnetic field $B=1T$ and the Fermi energy $\epsilon_F=-20 $ meV.  
}
\label{Fig-bi-freq}
\end{figure}

\section{Bilayer graphene and trigonal warping}

\subsection{Effective Hamiltonian}

AB-stacked bilayer graphene comprises two graphene sheets 
coupled by interlayer hoppings as shown in Fig.\ref{bilayer-lattice}, where 
B sublattice in the top sheet is located just above A sublattice of the 
bottom sheet (Bernal stacking).  
A unit cell thus contains four carbon atoms, i.e., 
A1 and B1 in the top layer and A2 and B2 in the bottom.  
Hoppings are 
the nearest-neighbor hopping within each layer ($\gamma_0$), 
the vertical interlayer hopping between B1 and A2 ($\gamma_1$).  
An oblique interlayer hopping between A1 and B2 ($\gamma_3$) 
causes the trigonal warping of the band dispersion. 
The magnitudes of these
are estimated in Ref.\cite{graphite-abinitio,dresselhaus-graphite02}
as $\gamma_0= 3.2$ eV,  $\gamma_1= 0.39$ eV,  $\gamma_3= 0.32$ eV.
We shall adopt these values hereafter.
There is another interlayer hopping parameters, $\gamma_4= 0.044$ eV, 
connecting A1 (B1) and A2 (B2), 
which introduces  a small electron-hole asymmetry to the band structure.
\cite{mcclure-gamma4-1,mcclure-gamma4-2}

In a basis with components, $\psi_{A1} , \psi_{B1},\psi_{A2} , \psi_{B2}$, 
on the four sites, 
the Hamiltonian for the bilayer graphene is given as
\cite{mccann-falko,guinea-stacks06,koshino2006transport}
\begin{equation}
H_{\rm AB}=
\begin{pmatrix}
0 &v \pi^\dagger & -v_4 \pi^\dagger & v_3 \pi \\
v \pi  &0& \gamma_1& -v_4 \pi^\dagger \\ 
-v_4 \pi & \gamma_1&0 & v \pi^\dagger \\ 
v_3 \pi^\dagger & -v_4 \pi & v \pi  &0 \\ 

\end{pmatrix}
,
\label{H-bi2}
\end{equation}
where $\pi=\xi \pi_x+i \pi_y$, $\pi^\dagger=\xi \pi_x-i \pi_y$,
$\Vec{\pi} = \Vec{p}+e\Vec{A}$, with 
$\Vec{A}$ being the vector potential arising from the 
applied magnetic field,
and the valley index $\xi=\pm 1$ for $K_\pm$ points.
Here $v=\sqrt{3} a \gamma_0/(2 \hbar)$ is the 
band velocity for monolayer graphene,
$a \approx 0.246$ nm the distance between the nearest $A$ sites,
and $v_3=\sqrt{3} a \gamma_3/(2 \hbar)$, $v_4=\sqrt{3} a \gamma_4/(2 \hbar)$ velocities 
related to $\gamma_3$, $\gamma_4$.

 The low-energy physics of bilayer graphene
for a region where the eigenenergy $\varepsilon$ is much smaller than 
the interlayer hopping $\gamma_1$, 
 is captured by a 2 $\times$ 2 Hamiltonian
in a basis of A1/B2 carbon sites,
with a perturbation with respect to $\varepsilon/\gamma_1$,
which is valid for energy scale $|\varepsilon| \ll \gamma_1=0.39eV$,
 as \cite{mccann-falko,koshino2006transport,koshino-LL-2011}
\begin{equation}
H^{\rm (eff)}_{\rm AB}
=\frac{1}{2m} 
\begin{pmatrix}
0 &(\pi^\dagger)^2 \\
\pi^2  &0 
\end{pmatrix}
+
v_3
\begin{pmatrix}
0 &\pi \\
\pi^\dagger &0 
\end{pmatrix}
+\frac{2v v_4}{\gamma_1}
\begin{pmatrix}
\pi^\dagger \pi & 0 \\
0 & \pi \pi^\dagger
\end{pmatrix}
,
\label{H-bilayer}
\end{equation}
with an effective mass $m=\gamma_1/(2 v^2)$.  
In the absence of magnetic fields,
the first term on the right-hand side 
gives a pair of parabolic bands $E = \pm p^2/(2m)$,
while the second term coming from $\gamma_3$
causes the trigonal warping in the band dispersion,
and  the third term from $v_4$ produces a
weak electron-hole asymmetry by adding the band energy
$2vv_4p^2/\gamma_1$ 
in both conduction and valence bands.
In the low-energy region, 
two touching parabolae are reformed into four
Dirac cones as shown in Fig.\ref{bilayer-lattice}(b),
where the Lifshitz transition (separation of the Fermi surface) occurs at
\begin{equation}
E_{\rm{Lifshitz}}= \frac{1}{2} m v_3^2  \sim 1\, \mbox{meV} . 
\end{equation}

The Landau level spectrum in a uniform magnetic field 
$\Vec{B} = {\rm rot}\Vec{A}$ may be found with 
the relation
$(\pi,\pi^\dagger) = (\sqrt{2}\hbar/\ell) (a^\dagger,a)$
for $K_+$, or $(\pi,\pi^\dagger) = (\sqrt{2}\hbar/\ell) (a,a^\dagger)$
for $K_-$.
Here $a^\dagger$ and $a$ are raising and lowering operators, respectively,
which operate on the Landau-level wave function $\phi_n$ as
$a \phi_n = \sqrt{n}\phi_{n-1}, 
a^\dagger \phi_n = \sqrt{n+1}\phi_{n+1}$, 
while $\ell = \sqrt{\hbar/(eB)}$ is the magnetic length.
If we neglect the $v_3$ and $v_4$ terms in Eqn.\ref{H-bilayer},
the eigenenergies are given by\cite{mccann-falko}
\begin{eqnarray}
&& \varepsilon_{n,s}= s \hbar \omega_c \sqrt{n(n+1)},
\end{eqnarray}
where states are labeled by the Landau index $n = -1,0,1,\cdots$,
and the band index $s = \pm$ (defined for $n\ge1$)
labeling the conduction ($s=+$) and valence  ($s=-$) bands.
The cyclotron frequency $\omega_c$ is 
the same as defined for a two-dimensional electron gas 
(2DEG),
\begin{eqnarray}
\hbar\omega_c&=& \frac{\hbar eB}{m} 
\simeq 3.5 \left( \frac{B}{1T}\right) \mbox{ meV}.
\end{eqnarray}
Two zero-energy Landau levels (LLs) appear ($n=0,-1$) 
at each valley,
while for large values of $n$ the LLs tend to be 
proportional to $n$ as in a 2DEG.  
The associated eigenstates for $K_+$ point are
\begin{eqnarray}
&&\psi_{n,s}=
C_n\begin{pmatrix}
\phi_{n-1} \\
s \phi_{n+1} \\ 
\end{pmatrix},
\nonumber \\
&& C_n = \left\{
\begin{array}{cc}
1 & (n\leq 0)\\ 
1/\sqrt{2} & (n\geq 1)
\end{array}
\right.,
\label{eqn_wf}
\end{eqnarray}
where we defined $\phi_n = 0$ for $n<0$.
The wavefunction at $K_-$ is obtained by
interchanging the first and second components 
in Eqn.\ref{eqn_wf}.

The $v_3$ term in Eqn. \ref{H-bilayer}
causes a hybridization between Landau states $\psi_{n,s}$ and
$\psi_{n+3m,s'}$,
 while $v_4$ contributes to the energy shift of the Landau levels.
 The energies and eigenfunctions 
for high-energy Landau levels with $|E| \gg E_{\rm Lifshitz}$ 
in $\hbar \omega_c \gg \hbar v_3/ \ell$ ($B>1$T),
are expressed in a lowest-order perturbation in $v_3$ and $v_4$ as
\begin{eqnarray}
&&
\tilde{\varepsilon}_{n,s} \simeq
\varepsilon_{n,s}+ \frac{2v v_4}{\gamma_1}\left(n+\frac{1}{2}\right),
\label{eq_energy_shift}
\\
&&\tilde{\psi}_{n,s} \simeq
\psi_{n,s}
+\sum_{s'=\pm}
\frac{\hbar v_3}{\sqrt 2 \ell} 
\left( 
\frac{s'\sqrt{n-1}\psi_{n-3,s'}
}{\varepsilon_{n,s} - \varepsilon_{n-3,s'}} 
+
\right.
\nonumber\\
&& \qquad\qquad\qquad\qquad
\left.
\frac{s\sqrt{n+2}\psi_{n+3,s'}
}{\varepsilon_{n,s} - \varepsilon_{n+3,s'}} 
\right).
\end{eqnarray}
In the low-energy region, $|E| < E_{\rm Lifshitz}$, 
on the other hand, 
the spectrum is reconstructed into
monolayer-like Landau levels 
as $\varepsilon_N={\rm sgn}(N)\sqrt{3N}\hbar \omega_{\rm{trig}}$
for the center cone and 
$\varepsilon_N={\rm sgn}(N)\sqrt{N}\hbar \omega_{\rm{trig}}$
for three off-center cones \cite{mccann-falko}, where 
\begin{equation}
\hbar \omega_{\rm{trig}}=v_3\sqrt{2\hbar eB}
\simeq 3.7 \sqrt{\frac{B}{1{\rm T}}}\mbox{ meV}.
\end{equation}
The first excited level $N=1$ in this series appears 
in small enough magnetic fields such that
$\hbar\omega_{\rm trig} < E_{\rm Lifshitz}$,
which amounts to $B < 0.08$T.

\subsection{Optical selection rules}

The optical longitudinal ($\sigma_{xx}(\omega)$) and Hall  ($\sigma_{xy}(\omega)$) conductivities are evaluated from the Kubo formula  as 
\begin{equation}
\sigma_{\alpha \beta}(\omega) =
 \frac{\hbar}{iL^2} \sum_{ab} j_\alpha^{ab} j_\beta^{ba} 
\frac{f(\epsilon_b) - f(\epsilon_a)}{\epsilon_b-\epsilon_a}
\frac{1}{\epsilon_b-\epsilon_a-\hbar\omega-i\eta},
\label{kuboformula}
\end{equation}
where $f(\varepsilon)$ is the Fermi distribution, 
$\epsilon_a$ the energy of the eigenstate $|a\rangle$, 
$j_\alpha^{ab} = \langle a | j_\alpha | b \rangle$
the matrix element of
the current operator $\Vec{j} = \partial H/\partial \Vec{A}$,
and $\eta$ a small energy 
cutoff for a stability of the calculation which we 
set to 0.1 meV. 

In multilayer graphene the optical selection rule 
is of crucial interest, which 
just reflects the 
current matrices, $j_x^{ab} j_y^{ba}$, appearing in the Kubo 
formula.  
Let us consider in the effective $2\times 2$ 
Hamiltonian Eqn.\ref{H-bilayer}, for which 
the current matrices read 
\begin{eqnarray}
j_x &=& \frac{\xi}{m} 
\begin{pmatrix}
0 &\pi^\dagger \\
\pi  &0 
\end{pmatrix}
+
\xi v_3
\begin{pmatrix}
0 & 1 \\
1 &0 
\end{pmatrix}
+
\frac{2v v_4}{\gamma_1}
\begin{pmatrix}
2\pi_x & 0 \\
0 & 2\pi_x
\end{pmatrix}
, 
\nonumber
\\
j_y &=& \frac{1}{m} 
\begin{pmatrix}
0 & -i \pi^\dagger \\
i \pi  &0 
\end{pmatrix}
+
v_3
\begin{pmatrix}
0 & i \\
-i &0 
\end{pmatrix}
+
\frac{2v v_4}{\gamma_1}
\begin{pmatrix}
2\pi_y & 0 \\
0 & 2\pi_y
\end{pmatrix}
.
\nonumber\\
\label{eq_current_operators}
\end{eqnarray}
The first term on the right-hand side of each equation 
is responsible for $n \leftrightarrow n\pm 1$ transitions.
When we express $a =(n,s)$ and $b =(n+1,s')$
in the Kubo formula (Eqn.\ref{kuboformula}),
the corresponding contribution is
\begin{equation}
j_x^{n,n+1} j_y^{n+1,n}=
- i (C_n C_{n+1})^2\frac{2 \hbar^2}{m^2 \ell^2}(n+1), 
\label{eq_jxjy_main}
 \end{equation}
where we have dropped the band indeces $s,s'$, since 
it is independent of their combination.

In addition, 
the second term in the current operators in Eqn.\
\ref{eq_current_operators},
and also the hybridization between $n$ and $n+3m$ in 
the eigenstates \cite{abergel-falko}, 
 give rise to additional transition series, 
\begin{eqnarray}
&& n \leftrightarrow n+1+3m,
\label{eq_sat1}\\
&& n \leftrightarrow n+2+3m,
\label{eq_sat2}
\end{eqnarray}
with all the combinations of $s$ and $s'$.
A similar selection rule was previously found for graphite
\cite{Falkovsky-graphite11,orlita-graphite12}.
In contrast, the third term including $v_4$ 
in Eqn.\ \ref{eq_current_operators}
contributes to the $n \leftrightarrow n\pm 1$ transistions
as the first term,
so that it does not cause any additional transitions
but adds a small correction to Eqn.\ \ref{eq_jxjy_main}.

For the high-energy Landau levels with $|E| \gg E_{\rm Lifshitz}$,
the contribution to $\sigma_{xy}(\omega)$
in the resonances of Eqns.\ (\ref{eq_sat1}) and (\ref{eq_sat2})
becomes of the order of $v_3^{2m}$ and $v_3^{2(m+1)}$, respectively.
In the lowest-order perturbation in $v_3$, the transitions 
are respectively 
allowed for $n \leftrightarrow n+4$ and $ n \leftrightarrow n+2$,
where the current matrices are respectively given by
\begin{eqnarray}
&&j_x^{n,n+4} j_y^{n+4,n} \approx
-i\left(\frac{v_3}{6}\right)^2,
\nonumber\\
&&j_x^{n,n+2} j_y^{n+2,n} \approx
i\left(\frac{v_3}{6}\right)^2.
\label{eq_jxjy_satellite}
\end{eqnarray}
Since the weights of these additional resonances have no dependence on the Landau index $n$ while the $n \leftrightarrow n+1$ resonance depends linearly on $n$, 
the resonances arising from the trigonal warping should be relatively 
prominent in a low Fermi energy ($\epsilon_F$) region.
We also note that the factor $j_x^{ab} j_y^{ba}$
may have positive or negative sign depending on the transition sequence.
This property is specific to $\sigma_{xy}$,
since the corresponding factors in $\sigma_{xx}$ 
 is $|j_x^{ab}|^2$, which naturally have no information about the sign of the resonances.

\subsection{Numerical results}

We start with calculating 
the optical Hall conductivity using Kubo formula Eqn.\ref{kuboformula} for the bilayer system using the effective Hamiltonian (Eqn.\ref{H-bilayer})
without the trigonal warping term. 
The result for the optical longitudinal and Hall conductivities in Fig.\ref{wo-warping}(a,b) shows that
an intra-band transition $(n,s) \leftrightarrow (n+1,s)$ occurs around 
\begin{equation}
 \hbar \omega_{\rm  intra} \sim  \hbar \omega_c.
\end{equation}
This is natural, since LLs are almost equally spaced, 
unlike in the monolayer case where LL energy $\propto \sqrt{n}$ is not equally separated.  On top of this, 
there are inter-band transitions across the band-touching point, with a selection rule $(n,\pm) \leftrightarrow (n+1,\mp)$.  
Thus the inter-band transition energy is 
\begin{equation}
\hbar\omega_{\rm inter} \simeq 2 |\epsilon_F|
\end{equation}
for large enough $n$.

As stressed above, an important difference between $\sigma_{xx}$ and $\sigma_{xy}$
is that resonance factor $j_\alpha^{ab} j_\beta^{ba}$  
is always positive for $\sigma_{xx}$, while
it may have both signs for $\sigma_{xy}$.
This puts a clear distinction for inter-band resonances of optical longitudinal and Hall conductivities.
For $\sigma_{xx}$, resonances $(n+1, -) \to (n,+)$ and 
$(n,-) \to (n+1,+)$ add up, and 
inter-band resonances occur over an entire region of Fermi energy
between negative $(n+1,-)$ and positive $(n+1,+)$ LL energies.  
By contrast, for $\sigma_{xy}$, resonances, 
$(n+1,-) \to (n,+)$ and $(n,-) \to (n+1,+)$ have 
opposites signs in the resonance factor, so that they 
cancel with each other for a region of Fermi energy between $(n,-)$ and $(n,+)$ LLs.  
So the inter-band resonance for $\sigma_{xy}$ is peaked at 
$\epsilon_{n+1,-} < \epsilon_F <\epsilon_{n,-} $ and
$\epsilon_{n,+} < \epsilon_F <\epsilon_{n+1,+} $ with opposite signs.


Now we examine the effect of the trigonal warping on the optical
responses. 
We go back to the original $4\times 4$ Hamiltonian (Eqn.\ref{H-bi2})
including all the hopping terms in order to get 
quantitatively accurate results for all the energy regions.
We have diagonalized the Hamiltonian
with a basis set spanned by a finite number of 
Landau functions $\phi_n$ ($0\le n<100$) 
to calculate the dynamical conductivities with
the formula Eqn.\ref{kuboformula}.
In Fig.\ref{Fig-bi2} we show a result 
for bilayer graphene in a magnetic field $B=1$T.
A behavior of LLs with magnetic fields (Landau fan diagram )
is indicated in Fig.\ref{Fig-bi2}(b).
At $B=1$T, the monolayer-like LLs in low-energy Dirac cones are not seen
as the LL spacing already exceeds $E_{\rm Lifshitz}$.

Figures \ref{Fig-bi2}(c,d)  are  $\sigma_{xx}(\epsilon_F,\omega)$ 
and $\sigma_{xy}(\epsilon_F,\omega)$, respectively, 
plotted against the Fermi energy $\epsilon_F$ and the frequency $\omega$.   
While the result becomes complicated,
we can make an identification of the resonance structure in 
Fig.\ \ref{Fig-bi2}(e), where we have extracted,
from the Landau level spectrum (Fig.\ref{Fig-bi2}(b)) at 1T, the expected 
positions of allowed resonances, as argued above.
There we can actually observe
the satellite resonances of $ n \leftrightarrow n+2$ and
$ n \leftrightarrow n+4$, which are absent in purely parabolic bilayer.
 It should be noted that, even though 
it is hard to directly observe the Landau levels below
the Lifshitz transition due to the tiny $E_{\rm{Lifshitz}}$,
the satellite transitions in a region outside of
$E_{\rm{Lifshitz}}$ should be observable as a manifestation of the
trigonal warping effect.

For $\sigma_{xy}(\omega)$,
the weight of the satellite peaks is constant as expected from
the current matrices of Eqn.\ \ref{eq_jxjy_satellite},
while that for the original peak ($ n \leftrightarrow n+1 $) 
depends on the Fermi energy through index $n$, as
$j_x^{n,n+1} j_y^{n+1,n} \simeq -i n \hbar^2/(2\ell^2 m^2) \simeq -i \epsilon_F/(2m)$.
So the relative weight of satellite to original peak 
is larger for smaller Landau index $n$ or smaller Fermi energy.

To make the signs of the resonances clearer, Figure \ref{Fig-bi-freq} 
plots the optical Hall conductivity $\sigma_{xy}$ against the frequency $\omega$ at a fixed Fermi energy $\epsilon_F=20$ meV.  
We can see that, in addition to a large resonance $n \to n+1$ around 
$\hbar \omega_c = 3$ meV,
satellite resonances $n \to n+2$ and $n \to n+4$ are seen.
In the case of the longitudinal conductivity $\sigma_{xx}$ the resonance appears as a symmetrical peak, 
while a resonance in $\sigma_{xy}$ has an odd structure around a resonance frequency $\omega_c$.
Thus the sign in the resonance weight for $\sigma_{xy}$ determines the resonance shape, i.e., 
in accordance with the sign of $ j_x^{ab}j_y^{ba} / i $ in Eqs. \ref{eq_jxjy_main} and \ref{eq_jxjy_satellite},
the resonance asymmetry of $n \to n+2$ is opposite to those of $n \to n+1$ and $n \to n+4$.
These features can be utilized for an experimental identification of resonances in the Faraday rotation,
which is directly proportional to $\sigma_{xy}(\omega)$ with Eqn.\ref{faraday},
where the unit $e^2/h$ in $\sigma_{xy}(\omega)$ is related to $2\alpha/(n_0+n_s)$ rad in Faraday rotation with the fine structure constant $\alpha \simeq 1/137$.

The resonance frequency for intra-band transition within the conduction band is larger than those within the valence band, which is a consequence of 
an electron-hole asymmetry due to $\gamma_4$ term.  
A deviation in the  cyclotron mass for electron and hole bands prevents a complete cancellation between $(n,-) \to (n+1,+)$ and $(n+1,-) \to (n,+)$ transitions, which result in small interband transitions in a wide region of Fermi energy.


\section{Trilayer graphene}

\begin{figure}[tb]
\begin{center}
\includegraphics[width=0.9\linewidth]{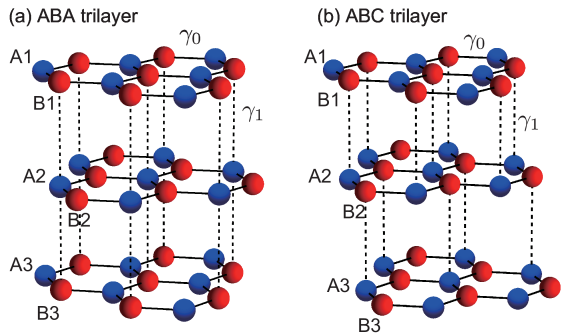}
\end{center}
\caption{
Lattice structure of trilayer graphene with (a) ABA stacking and (b) ABC stacking.
}
\label{trilayer-lattice}
\end{figure}

Now let us move on to optical responses of trilayer graphene QHE systems.
Trilayer graphene occurs in two different stacking orders, 
i.e., ABA and ABC stacking orders as shown in Fig.\ref{trilayer-lattice}.
In both ABA and ABC trilayers,
the spatial arrangement of the two successive layers 
is the similar to bilayer graphene, 
while the relation between the first and third layers 
is different between two.
The variation in the stacking order results in
totally different electronic structures, and here we 
show that this can be dramatically reflected 
in optical responses.

\subsection{ABA-stacked trilayer}

For ABA stacked trilayer,
the effective Hamiltonian around $K_+$/$K_-$ points 
is given by a $6\times 6$ matrix 
(the dimension being 2 sublattices $\times$ 3 layers) 
as \cite{guinea-stacks06,partoens2006graphene,lu2006influence,koshino-ando07}
\begin{equation}
H_{\rm ABA}=
\begin{pmatrix}
0 &v \pi^\dagger & -v_4 \pi^\dagger &v_3 \pi &\gamma_2/2&0\\
v \pi  &\Delta'& \gamma_1& -v_4 \pi^\dagger &0 & \gamma_5/2 \\ 
-v_4 \pi& \gamma_1&\Delta' & v \pi^\dagger &-v_4 \pi & \gamma_1\\ 
v_3 \pi^\dagger&-v_4 \pi& v \pi  &0 &v_3 \pi^\dagger&-v_4 \pi \\ 
\gamma_2/2&0& -v_4 \pi^\dagger&v_3 \pi &0& v \pi^\dagger  \\ 
0&\gamma_5/2& \gamma_1 &-v_4 \pi^\dagger& v \pi&\Delta'  
\end{pmatrix},
\end{equation}
where $\gamma_0$, $\gamma_1$ and $\gamma_3$
are the same hopping parameters as in bilayer, 
$\Delta'$ the on-site energy difference between
the atoms with and without vertical bond $\gamma_1$,
and $\gamma_2 (\gamma_5)$ the 
next-nearest interlayer hoppings between $A1$ and $A3$ 
($B1$ and $B3$).  
We adopt the values for bulk graphite, 
$\gamma_2$=-0.020eV, 
$\gamma_5$=0.038eV, 
$\Delta'$=0.050eV.
\cite{graphite-abinitio,dresselhaus-graphite02}

\begin{figure*}[tb]
\begin{center}
\includegraphics[width=0.95\linewidth]{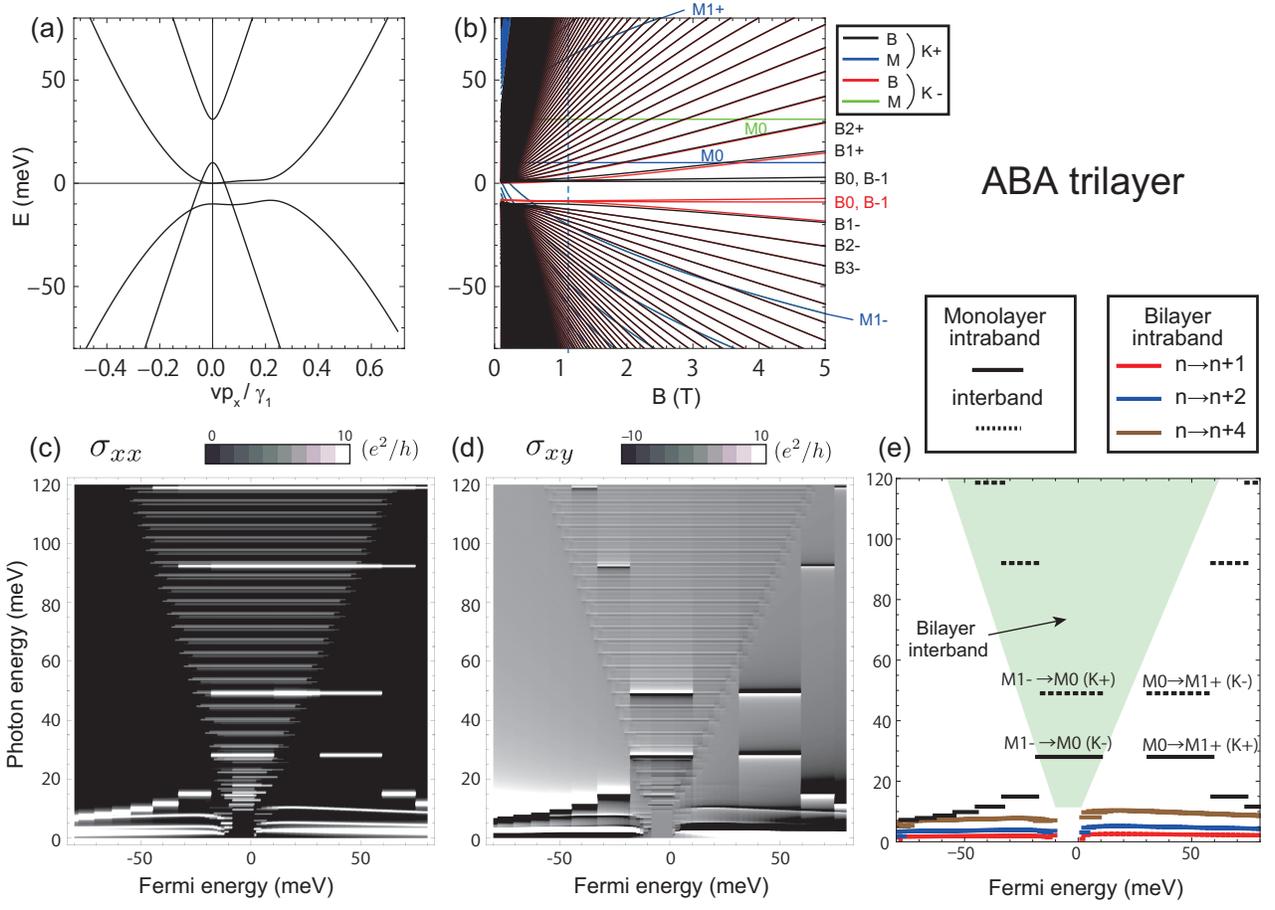}
\end{center}
\caption{
(a) Low-energy band structure, and (b) Landau levels against magnetic field for ABA-stacked trilayer graphene.
(c) Longitudinal $\sigma_{xx}(\epsilon_F,\omega)$ and (d) Hall 
$\sigma_{xy}(\epsilon_F,\omega)$ 
plotted against the Fermi energy $\epsilon_F$ and the frequency $\omega$
for a magnetic field $B=1T$ (a dashed line in (b)). 
(e) A diagram indicating allowed resonances in $\sigma_{xy}$.  
}
\label{Fig-aba}
\end{figure*}

With a unitary transformation 
\cite{koshino-ando07,koshino-ando08,koshino-aba-2009,koshino2009electronic,koshino-LL-2011},
this Hamiltonian is decomposed into two blocks as
\begin{equation}
\tilde{H}_{\rm ABA}=
\begin{pmatrix}
H_m & 0 \\
0 & H_b \\
\end{pmatrix}
.
\end{equation}
The first 2 by 2 block,
\begin{equation}
H_m=
\begin{pmatrix}
-\gamma_2/2 & v \pi^\dagger\\
v \pi & -\gamma_5/2 + \Delta'\\
\end{pmatrix}
,
\end{equation}
 corresponds to a massive Dirac Hamiltonian with a shift in Fermi energy,
while the other 4 by 4 block, 
\begin{equation}
H_b=
\begin{pmatrix}
\gamma_2/2 &v \pi^\dagger & -\sqrt{2} v_4 \pi^\dagger& \sqrt{2} v_3 \pi \\
v \pi  &\gamma_5/2+\Delta'& \sqrt{2}\gamma_1& -\sqrt{2} v_4 \pi^\dagger\\ 
-\sqrt{2} v_4 \pi & \sqrt{2}\gamma_1&\Delta' & v \pi^\dagger \\ 
\sqrt{2} v_3 \pi^\dagger & -\sqrt{2} v_4 \pi & v \pi  &0 \\ 
\end{pmatrix}
,
\end{equation}
corresponds to a gapped bilayer Hamiltonian with another shift in energy.
So the low-energy physics of ABA stacked trilayer graphene is effectively described as a superposition 
of gapped monolayer and gapped bilayer band contributions
as seen in the low-energy band structure in Fig.\ref{Fig-aba}(a).

In Fig.\ref{Fig-aba}(b), LLs are plotted with magnetic field B,
which are labeled 
with M (B) for monolayer (bilayer) blocks, 
Landau index $n$, and band index $\pm$ (for $n\ge 1$).
For moderate magnetic fields $B \sim 1$ T, 
the Landau level spacing $\propto \sqrt B$ 
for monolayer is much larger than that for bilayer $\propto B$.
Since bands are gapped, zero-energy LLs of monolayer and bilayer 
appear at the bottom (top) 
of the conduction (valence) bands for $K_+$ ($K_-$) valley.

We now look at the result for the optical longitudinal $\sigma_{xx}(\epsilon_F,\omega)$ and Hall conductivity $\sigma_{xy}(\epsilon_F,\omega)$
plotted against the Fermi energy $\epsilon_F$ and frequency $\omega$
for ABA stacked trilayer graphene QHE system in Figs.\ref{Fig-aba}(c,d).
We can discern contributions from monolayer-like Dirac LLs and 
from bilayer LLs, both of which exhibit intra-band and inter-band transitions.
Since Dirac cone is massive and M0 LL is situated at the  bottom of
conduction band for $K_+$ valley and the top of valence band for $K_-$,
 so that 
$M0 \to M1+$ resonance occurs at a lower energy than $M1- \to M0$ for $K_+$, and vice versa for $K_-$ valley.  
A cancellation of resonances in $\sigma_{xy}$, due to opposite signs in current matrices, 
occurs between $M1- \to M0$
for $K_+$ and  $M0 \to M1+$ for $K_-$ 
for a region of Fermi energy between $M0(K_+)$ and $M0(K_-)$,
while this is not the case with $\sigma_{xx}$.  
For bilayer contributions satellites appear due to the trigonal warping effect
as 
$(n,s) \leftrightarrow (n+1+3m,s')$ and $(n,s) \leftrightarrow (n+2+3m,s')$ 
as in optical responses for bilayer graphene (Fig.\ref{Fig-bi2}).

So the message here is the ac optical responses 
in the ABA-stacked trilayer graphene 
accommodates a curious mixture of contributions 
from an effective massive monolayer and from an effective gapped bilayer with the trigonal warping effect. 

\subsection{ABC-stacked trilayer}
If we turn to ABC stacked bilayer graphene,
the effective Hamiltonian around $K$ point is a $6\times 6$ matrix as 
\cite{guinea-stacks06,lu-2006-abc,koshino-abc-2009}
\begin{equation}
 H_{\rm ABC}=
\begin{pmatrix}
0 &v \pi^\dagger & -v_4 \pi^\dagger &v_3 \pi &0&\gamma_2/2\\
v \pi  &0& \gamma_1&-v_4 \pi^\dagger &0 & 0 \\ 
-v_4 \pi& \gamma_1&0 & v \pi^\dagger &-v_4 \pi^\dagger & v_3 \pi\\ 
v_3 \pi^\dagger&-v_4 \pi& v \pi  &0 &\gamma_1&-v_4 \pi^\dagger \\ 
0&0& -v_4 \pi&\gamma_1 &0 & v \pi^\dagger  \\ 
\gamma_2/2&0& v_3 \pi^\dagger&-v_4 \pi& v \pi&0  
\end{pmatrix}
.
\label{H-abc}
\end{equation}
We can derive a low-energy effective Hamiltonian 
as a $2\times 2$ matrix with basis for A1 and B3, where we 
eliminate the states coupled by $\gamma_1$.
As in the case of bilayer, 
a perturbation in $\varepsilon/\gamma_1$
gives the effective Hamiltonian for ABC trilayer graphene as
\cite{koshino-abc-2009},
\begin{equation}
 H^{\rm (eff)}_{\rm ABC}=
\frac{v^3}{\gamma_1^2}
\begin{pmatrix}
0 & (\pi^\dagger)^3\\
\pi^3  & 0
\end{pmatrix}
+
\left(
\frac{\gamma_2}{2}
-\frac{2v v_3 \Vec{\pi}^2}{\gamma_1}
\right)
\begin{pmatrix}
0 & 1 \\
1  & 0
\end{pmatrix},
\label{H-abc-eff}
\end{equation}
where $\Vec{\pi}^2=(\pi^\dagger \pi + \pi \pi^\dagger)/2$,
and we neglected $v_4$ term which gives the electron-hole asymmetry
in a similar manner to bilayer graphene.
In the absence of magnetic fields,
the first term gives a pair of cubic-dispersed 
bands touching at zero energy,
while the second term involving $\gamma_2$ and $v_3$
causes a trigonal warping in the band dispersion.
In a low-energy region, 
the cubic bands are reformed into three
Dirac cones at off center momenta located in $120^\circ$
symmetry around $K_\pm$ point.
The Lifshitz transition occurs at
\begin{equation}
E_{\rm{Lifshitz}} \approx \frac{\gamma_2}{2}  \sim 10\, \mbox{meV},
\end{equation}
which is an order of magnitude greater than in bilayer's.

If we first neglect $\gamma_2$ and $v_3$ and consider the cubic part alone in Eqn.\ \ref{H-abc-eff},
LLs are\cite{guinea-stacks06,koshino-abc-2009}
\begin{equation}
\varepsilon_{n,s}=s \hbar \omega_{\rm{ABC}} \sqrt{n(n+1)(n+2)}
,
\label{LL-abc}
\end{equation}
where $n \ge -2$ is the Landau index, 
$s=\pm$ (only for $n \geq 1$) the band index,
and
\begin{equation}
\hbar \omega_{\rm{ABC}}
=\frac{v^3}{\gamma_1^2} \left(\frac{\sqrt{2} \hbar}{\ell}\right)^3
=\frac{v^3}{\gamma_1^2}  (2 \hbar eB )^{\frac 3 2 }.
\end{equation}
A peculiar magnetic field dependence of cyclotron energy $\propto B^{\frac 3 2}$
for ABC trilayer graphene
implies a smaller LL 
spacing compared to the monolayer's LL $\propto B^{\frac 1 2}$ 
and bilayer's LL $\propto B$ for weak magnetic fields.  
The associated wavefunction at $K_+$ point is
\begin{equation}
\psi_{n,s}=
C_n \begin{pmatrix}
\phi_{n-1} \\
s \phi_{n+2}
\end{pmatrix},
\label{eq_LL_abc_wav}
\end{equation}
where $C_n$ is defined in Eqn.\ \ref{eqn_wf}.
The wavefunction at $K_-$ is obtained by
interchanging the first and second components 
in Eqn.\ \ref{eq_LL_abc_wav}.

Similar to bilayer, the trigonal warping effect 
due to $\gamma_2$ and $v_3$
hybridizes $\psi_{n,s}$ with $\psi_{n+3m,s'}$.
In the low-energy region $|E| < E_{\rm Lifshitz}$, 
the spectrum is reconstructed into
the monolayer-like Landau levels from small Dirac cones
as a series $\varepsilon_N={\rm sgn}(N)\sqrt{N}\hbar \omega_{\rm{trig}}$, 
where
\begin{eqnarray}
 \hbar \omega_{\rm{trig}}&=&
3\left|
\frac{\gamma_2}{2\gamma_1}
\right|^{2/3}
\left[
1+\frac{4v_3}{3v}
\left(
\frac{2\gamma_1}{\gamma_2}
\right)^{1/3}
\right]^{1/2}
v \sqrt{2\hbar eB}
\nonumber\\
&\simeq& 10 \sqrt{\frac{B}{1{\rm T}}}\mbox{ meV}.
\end{eqnarray}
Each monolayer-like level 
is three-fold degenerate reflecting the number of small Dirac cones.
For larger magnetic fields 
the degeneracy of the $N$-th level is lifted 
when $\varepsilon_N$ exceeds $E_{\rm Lifshitz}$,
and the three levels eventually connect to 
$n = 3N, 3N-1$ and $3N-2$ in the cubic band 
in Eqn.\ \ref{LL-abc}. 
The first excited level $N=1$ can appear when
$\hbar\omega_{\rm trig} < E_{\rm Lifshitz}$ or
$B <\sim 1$T, which is much larger than in bilayer graphene.

\begin{figure*}[tb]
\begin{center}
\includegraphics[width=0.95\linewidth]{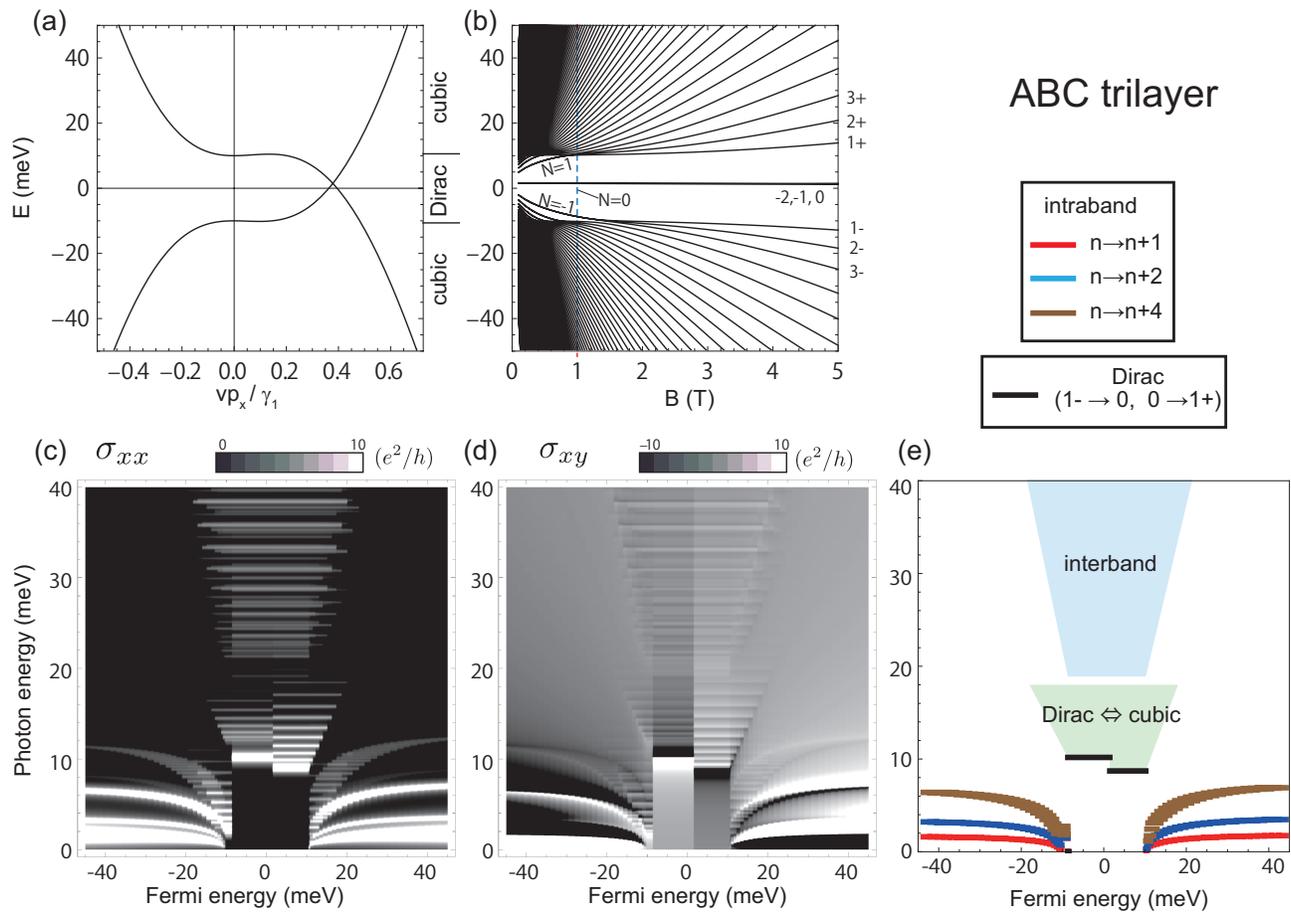}
\end{center}
\caption{
(a) Low-energy band structure, and (b) Landau levels against magnetic field for ABC-stacked trilayer graphene.
(c) Longitudinal $\sigma_{xx}(\epsilon_F,\omega)$ and (d) Hall $\sigma_{xy}(\epsilon_F,\omega)$ 
plotted against the Fermi energy $\epsilon_F$ and the frequency $\omega$
for a magnetic field $B=1T$ (a dashed line in (b)). 
(e) A diagram indicating allowed resonances in $\sigma_{xy}$.  
}
\label{Fig-abc}
\end{figure*}

Now we go back to the original $6\times 6$ Hamiltonian including all the hopping terms (Eqn.\ref{H-abc}) 
to discuss optical conductivities calculated with diagonalization 
and Kubo formula.  
In a low-energy region ($\sim$ 10 meV), Dirac cones appear 
due to trigonal warping effects as seen 
in the band dispersion (Fig.\ref{Fig-abc}(a)).
In the LL spectrum in Fig.\ref{Fig-abc}(b),
a Lifshitz transition is clearly identified at 
$E_{\rm Lifshitz}=10$meV, 
which separates Dirac LLs and ABC cubic LLs.

Figures \ref{Fig-abc}(c,d) show the optical longitudinal $\sigma_{xx}(\epsilon_F,\omega)$ and  Hall conductivities  $\sigma_{xy}(\epsilon_F,\omega)$
plotted against the Fermi energy $\epsilon_F$ and frequency $\omega$
for ABC stacked trilayer graphene QHE system with $B=1$T.  
At this particular magnetic field, there are three Landau levels
$N=-1,0,1$ inside each Dirac cone,
where the  $N =\pm 1$ levels are close to the Lifshitz transision.  
The higher levels are outside of the Dirac cones, 
and can be regarded to belong to the the cubic dispersion.
Accordingly we see cyclotron resonances 
between $1- \to 0$ and $0 \to 1+$ in $E_F < E_{\rm Lifshitz}$,
while outside we see transitions between cubic LLs $\propto B^{\frac 3 2}$ with much smaller level spacings 
than for Dirac LLs $\propto \sqrt{B}$.
In addition to resonances $ n \leftrightarrow n+1$,
 the trigonal warping again gives rise to satellite transitions 
$ n \leftrightarrow n+4$ and $ n \leftrightarrow n+2$ 
with opposite resonance weights.
There also emerge small resonances between Dirac $n=0$ LL and
cubic LLs across the Lifshitz transition energy.
Due to the ABC trilayer LLs arising from the 
cubic dispersion (Eqn.\ref{LL-abc})  the intra-band transition energies show behaviors $\propto n^{1/2}$ with Landau index $n$,
which is different from monolayer ($\propto n^{-1/2}$) or bilayer (constant),
while the inter-band transition energies are qualitatively similar to $\sim 2\epsilon_F$.

Thus an message in the case of ABC trilayer graphene 
is that the trigonal warping effect is an order of magnitude enhanced 
than in bilayer graphene,
so it will be experimentally more feasible to access the effects induced by the trigonal warping.

\section{Summary}
We have studied the optical Hall conductivity for bilayer and trilayer graphene.
In bilayer graphene the trigonal warping effect causes a Lifshitz transition with an emergence of four Dirac cones in a small energy scale $\sim 1 \mbox{ meV}$.
The trigonal warping is shown here to affect optical transitions as additional transitions arising from mixing of LLs whose Landau indices differ by 3 
and the opposite sign of resonance weights for transitions $n \leftrightarrow n+2$.
For trilayer graphene, which comes with two stacking orders,
we have shown that 
the optical conductivity of ABA in low-energy region comprises 
contributionf from of monolayer-like massive Dirac and gapped bilayer 
with an electron-hole asymmetry,
while ABC stacking has an order of magnitude larger trigonal warping effect,
exhibiting resonances of Dirac LLs inside the Lifshitz transition ($\sim 10$ meV),
and a sequence of resonances for the cubic bands with cyclotron energy $\propto B^{\frac 3 2}$.   This is predicted to be observable.

 There are several reports that the electron-electron interaction 
opens an band gap of a few meV at charge neutrality point
in bilayer and trilayer graphene \cite{velasco2012,bao-abc2011}.  
If we consider the effect of the
many-body induced gap in the effective models phenomenologically, this
would cause a valley splitting of $n=0$ LL and 
energy shift of resonances associated with low-lying levels, 
while the transitions associated to high LLs away from the
Dirac points, including those of satellite transitions, 
should not be much affected. 
We leave the calculation including those effects for future works.

There are various other problems.  
One is effects of disorder, which in this paper 
is treated as a phenomenological broadening of the energies, 
assumed to be small.  
However, the localization physics is important for understanding the quantum Hall physics, 
so that it should be interesting to treat disorder more accurately with methods such as diagonalization technique as done in Ref.\cite{morimoto-ac-scaling,morimoto-two-parameter-scaling}, and consider the localization effects on multilayer graphene in the quantum Hall regime.  
It would be another interesting problem to ask how various  types of disorder, chiral-symmetry preserving and non-preserving,
affect the optical responses in the multilayer graphene, 
since  it is known that the presence or absence of chiral symmetry 
greatly affects the behavior of zero energy LLs for the bilayer graphene \cite{kawarabayashi-chiral-bilayer}.

\section*{Acknowledgments}
HA acknowledges the financial support by Grants-in-Aid for Scientific Research, 
No.23340112 from JSPS.
MK acknowledges the financial supports by
Grants-in-Aid for Scientific Research, 
No.24740193 from JSPS,
and  by JST-EPSRC Japan-UK
Cooperative Programme Grant No. EP/H025804/1.

\bibliography{thesis,opthall,honeycomb,Publist,scaling,bilayer,koshino}

\begin{thebibliography}{42}
\expandafter\ifx\csname natexlab\endcsname\relax\def\natexlab#1{#1}\fi
\expandafter\ifx\csname bibnamefont\endcsname\relax
  \def\bibnamefont#1{#1}\fi
\expandafter\ifx\csname bibfnamefont\endcsname\relax
  \def\bibfnamefont#1{#1}\fi
\expandafter\ifx\csname citenamefont\endcsname\relax
  \def\citenamefont#1{#1}\fi
\expandafter\ifx\csname url\endcsname\relax
  \def\url#1{\texttt{#1}}\fi
\expandafter\ifx\csname urlprefix\endcsname\relax\def\urlprefix{URL }\fi
\providecommand{\bibinfo}[2]{#2}
\providecommand{\eprint}[2][]{\url{#2}}

\bibitem[{\citenamefont{Novoselov et~al.}(2005)\citenamefont{Novoselov, Geim,
  Morozov, Jiang, Katsnelson, Grigorieva, Dubonos, and Firsov}}]{Nov05}
\bibinfo{author}{\bibfnamefont{K.}~\bibnamefont{Novoselov}},
  \bibinfo{author}{\bibfnamefont{A.}~\bibnamefont{Geim}},
  \bibinfo{author}{\bibfnamefont{S.}~\bibnamefont{Morozov}},
  \bibinfo{author}{\bibfnamefont{D.}~\bibnamefont{Jiang}},
  \bibinfo{author}{\bibfnamefont{M.}~\bibnamefont{Katsnelson}},
  \bibinfo{author}{\bibfnamefont{I.}~\bibnamefont{Grigorieva}},
  \bibinfo{author}{\bibfnamefont{S.}~\bibnamefont{Dubonos}}, \bibnamefont{and}
  \bibinfo{author}{\bibfnamefont{A.}~\bibnamefont{Firsov}},
  \bibinfo{journal}{Nature} \textbf{\bibinfo{volume}{438}},
  \bibinfo{pages}{197} (\bibinfo{year}{2005}).

\bibitem[{\citenamefont{Zhang et~al.}(2005)\citenamefont{Zhang, Tan, Stormer,
  and Kim}}]{Zha05}
\bibinfo{author}{\bibfnamefont{Y.}~\bibnamefont{Zhang}},
  \bibinfo{author}{\bibfnamefont{Y.~W.} \bibnamefont{Tan}},
  \bibinfo{author}{\bibfnamefont{H.~L.} \bibnamefont{Stormer}},
  \bibnamefont{and} \bibinfo{author}{\bibfnamefont{P.}~\bibnamefont{Kim}},
  \bibinfo{journal}{Nature} \textbf{\bibinfo{volume}{438}},
  \bibinfo{pages}{201} (\bibinfo{year}{2005}).

\bibitem[{\citenamefont{Gusynin et~al.}(2007)\citenamefont{Gusynin, Sharapov,
  and Carbotte}}]{gusynin-sxy}
\bibinfo{author}{\bibfnamefont{V.~P.} \bibnamefont{Gusynin}},
  \bibinfo{author}{\bibfnamefont{S.~G.} \bibnamefont{Sharapov}},
  \bibnamefont{and} \bibinfo{author}{\bibfnamefont{J.~P.}
  \bibnamefont{Carbotte}}, \bibinfo{journal}{J. Phys.: Condens. Matter}
  \textbf{\bibinfo{volume}{19}}, \bibinfo{pages}{026222}
  (\bibinfo{year}{2007}).

\bibitem[{\citenamefont{Morimoto et~al.}(2009)\citenamefont{Morimoto, Hatsugai,
  and Aoki}}]{morimoto-opthall}
\bibinfo{author}{\bibfnamefont{T.}~\bibnamefont{Morimoto}},
  \bibinfo{author}{\bibfnamefont{Y.}~\bibnamefont{Hatsugai}}, \bibnamefont{and}
  \bibinfo{author}{\bibfnamefont{H.}~\bibnamefont{Aoki}},
  \bibinfo{journal}{Phys. Rev. Lett.} \textbf{\bibinfo{volume}{103}},
  \bibinfo{pages}{116803} (\bibinfo{year}{2009}).

\bibitem[{\citenamefont{Fialkovsky and Vassilevich}(2009)}]{Fialkovsky09}
\bibinfo{author}{\bibfnamefont{I.~V.} \bibnamefont{Fialkovsky}}
  \bibnamefont{and} \bibinfo{author}{\bibfnamefont{D.~V.}
  \bibnamefont{Vassilevich}}, \bibinfo{journal}{J. Phys. A: Math. Theor.}
  \textbf{\bibinfo{volume}{42}}, \bibinfo{pages}{442001}
  (\bibinfo{year}{2009}).

\bibitem[{\citenamefont{Falkovsky}(2011)}]{Falkovsky-graphite11}
\bibinfo{author}{\bibfnamefont{L.~A.} \bibnamefont{Falkovsky}},
  \bibinfo{journal}{Phys. Rev. B} \textbf{\bibinfo{volume}{84}},
  \bibinfo{pages}{115414} (\bibinfo{year}{2011}).

\bibitem[{\citenamefont{Tse and MacDonald}(2010)}]{tse-macdonald10}
\bibinfo{author}{\bibfnamefont{W.-K.} \bibnamefont{Tse}} \bibnamefont{and}
  \bibinfo{author}{\bibfnamefont{A.~H.} \bibnamefont{MacDonald}},
  \bibinfo{journal}{Phys. Rev. B} \textbf{\bibinfo{volume}{82}},
  \bibinfo{pages}{161104} (\bibinfo{year}{2010}).

\bibitem[{\citenamefont{O'Connell and Wallace}(1982)}]{oconnell82}
\bibinfo{author}{\bibfnamefont{R.~F.} \bibnamefont{O'Connell}}
  \bibnamefont{and} \bibinfo{author}{\bibfnamefont{G.}~\bibnamefont{Wallace}},
  \bibinfo{journal}{Phys. Rev. B} \textbf{\bibinfo{volume}{26}},
  \bibinfo{pages}{2231} (\bibinfo{year}{1982}).

\bibitem[{\citenamefont{Ikebe et~al.}(2010)\citenamefont{Ikebe, Morimoto,
  Masutomi, Okamoto, Aoki, and Shimano}}]{ikebe-THz10}
\bibinfo{author}{\bibfnamefont{Y.}~\bibnamefont{Ikebe}},
  \bibinfo{author}{\bibfnamefont{T.}~\bibnamefont{Morimoto}},
  \bibinfo{author}{\bibfnamefont{R.}~\bibnamefont{Masutomi}},
  \bibinfo{author}{\bibfnamefont{T.}~\bibnamefont{Okamoto}},
  \bibinfo{author}{\bibfnamefont{H.}~\bibnamefont{Aoki}}, \bibnamefont{and}
  \bibinfo{author}{\bibfnamefont{R.}~\bibnamefont{Shimano}},
  \bibinfo{journal}{Phys. Rev. Lett.} \textbf{\bibinfo{volume}{104}},
  \bibinfo{pages}{256802} (\bibinfo{year}{2010}).

\bibitem[{\citenamefont{Crassee et~al.}(2011)\citenamefont{Crassee, Levallois,
  Walter, Ostler, Bostwick, Rotenberg, Seyller, Van Der~Marel, and
  Kuzmenko}}]{crassee2010giant}
\bibinfo{author}{\bibfnamefont{I.}~\bibnamefont{Crassee}},
  \bibinfo{author}{\bibfnamefont{J.}~\bibnamefont{Levallois}},
  \bibinfo{author}{\bibfnamefont{A.}~\bibnamefont{Walter}},
  \bibinfo{author}{\bibfnamefont{M.}~\bibnamefont{Ostler}},
  \bibinfo{author}{\bibfnamefont{A.}~\bibnamefont{Bostwick}},
  \bibinfo{author}{\bibfnamefont{E.}~\bibnamefont{Rotenberg}},
  \bibinfo{author}{\bibfnamefont{T.}~\bibnamefont{Seyller}},
  \bibinfo{author}{\bibfnamefont{D.}~\bibnamefont{Van Der~Marel}},
  \bibnamefont{and} \bibinfo{author}{\bibfnamefont{A.}~\bibnamefont{Kuzmenko}},
  \bibinfo{journal}{Nature Physics} \textbf{\bibinfo{volume}{7}},
  \bibinfo{pages}{48} (\bibinfo{year}{2011}).

\bibitem[{\citenamefont{Novoselov et~al.}(2006)\citenamefont{Novoselov, McCann,
  Morozov, Falfko, Katsnelson, Zeitler, Jiang, Schedin, and
  Geim}}]{nov-bilayer}
\bibinfo{author}{\bibfnamefont{K.}~\bibnamefont{Novoselov}},
  \bibinfo{author}{\bibfnamefont{E.}~\bibnamefont{McCann}},
  \bibinfo{author}{\bibfnamefont{S.}~\bibnamefont{Morozov}},
  \bibinfo{author}{\bibfnamefont{V.}~\bibnamefont{Falfko}},
  \bibinfo{author}{\bibfnamefont{M.}~\bibnamefont{Katsnelson}},
  \bibinfo{author}{\bibfnamefont{U.}~\bibnamefont{Zeitler}},
  \bibinfo{author}{\bibfnamefont{D.}~\bibnamefont{Jiang}},
  \bibinfo{author}{\bibfnamefont{F.}~\bibnamefont{Schedin}}, \bibnamefont{and}
  \bibinfo{author}{\bibfnamefont{A.}~\bibnamefont{Geim}},
  \bibinfo{journal}{Nature Phys.} \textbf{\bibinfo{volume}{2}},
  \bibinfo{pages}{177} (\bibinfo{year}{2006}).

\bibitem[{\citenamefont{McCann and Falfko}(2006)}]{mccann-falko}
\bibinfo{author}{\bibfnamefont{E.}~\bibnamefont{McCann}} \bibnamefont{and}
  \bibinfo{author}{\bibfnamefont{V.}~\bibnamefont{Falfko}},
  \bibinfo{journal}{Phys. Rev. Lett.} \textbf{\bibinfo{volume}{96}},
  \bibinfo{pages}{86805} (\bibinfo{year}{2006}).

\bibitem[{\citenamefont{Abergel and Fal'ko}(2007)}]{abergel-falko}
\bibinfo{author}{\bibfnamefont{D.~S.~L.} \bibnamefont{Abergel}}
  \bibnamefont{and} \bibinfo{author}{\bibfnamefont{V.~I.}
  \bibnamefont{Fal'ko}}, \bibinfo{journal}{Phys. Rev. B}
  \textbf{\bibinfo{volume}{75}}, \bibinfo{pages}{155430}
  (\bibinfo{year}{2007}).

\bibitem[{\citenamefont{Nicol and Carbotte}(2008)}]{carbotte-sxx-bilayer}
\bibinfo{author}{\bibfnamefont{E.~J.} \bibnamefont{Nicol}} \bibnamefont{and}
  \bibinfo{author}{\bibfnamefont{J.~P.} \bibnamefont{Carbotte}},
  \bibinfo{journal}{Phys. Rev. B} \textbf{\bibinfo{volume}{77}},
  \bibinfo{pages}{155409} (\bibinfo{year}{2008}).

\bibitem[{\citenamefont{Zhang et~al.}(2008)\citenamefont{Zhang, Li, Basov,
  Fogler, Hao, and Martin}}]{determining-bilayer-band}
\bibinfo{author}{\bibfnamefont{L.~M.} \bibnamefont{Zhang}},
  \bibinfo{author}{\bibfnamefont{Z.~Q.} \bibnamefont{Li}},
  \bibinfo{author}{\bibfnamefont{D.~N.} \bibnamefont{Basov}},
  \bibinfo{author}{\bibfnamefont{M.~M.} \bibnamefont{Fogler}},
  \bibinfo{author}{\bibfnamefont{Z.}~\bibnamefont{Hao}}, \bibnamefont{and}
  \bibinfo{author}{\bibfnamefont{M.~C.} \bibnamefont{Martin}},
  \bibinfo{journal}{Phys. Rev. B} \textbf{\bibinfo{volume}{78}},
  \bibinfo{pages}{235408} (\bibinfo{year}{2008}).

\bibitem[{\citenamefont{Li et~al.}(2009)\citenamefont{Li, Henriksen, Jiang,
  Hao, Martin, Kim, Stormer, and Basov}}]{band-asymmetry-bilayer}
\bibinfo{author}{\bibfnamefont{Z.~Q.} \bibnamefont{Li}},
  \bibinfo{author}{\bibfnamefont{E.~A.} \bibnamefont{Henriksen}},
  \bibinfo{author}{\bibfnamefont{Z.}~\bibnamefont{Jiang}},
  \bibinfo{author}{\bibfnamefont{Z.}~\bibnamefont{Hao}},
  \bibinfo{author}{\bibfnamefont{M.~C.} \bibnamefont{Martin}},
  \bibinfo{author}{\bibfnamefont{P.}~\bibnamefont{Kim}},
  \bibinfo{author}{\bibfnamefont{H.~L.} \bibnamefont{Stormer}},
  \bibnamefont{and} \bibinfo{author}{\bibfnamefont{D.~N.} \bibnamefont{Basov}},
  \bibinfo{journal}{Phys. Rev. Lett.} \textbf{\bibinfo{volume}{102}},
  \bibinfo{pages}{037403} (\bibinfo{year}{2009}).

\bibitem[{\citenamefont{Guinea et~al.}(2006)\citenamefont{Guinea, Castro~Neto,
  and Peres}}]{guinea-stacks06}
\bibinfo{author}{\bibfnamefont{F.}~\bibnamefont{Guinea}},
  \bibinfo{author}{\bibfnamefont{A.~H.} \bibnamefont{Castro~Neto}},
  \bibnamefont{and} \bibinfo{author}{\bibfnamefont{N.~M.~R.}
  \bibnamefont{Peres}}, \bibinfo{journal}{Phys. Rev. B}
  \textbf{\bibinfo{volume}{73}}, \bibinfo{pages}{245426}
  (\bibinfo{year}{2006}).

\bibitem[{\citenamefont{Latil and Henrard}(2006)}]{latil-2006}
\bibinfo{author}{\bibfnamefont{S.}~\bibnamefont{Latil}} \bibnamefont{and}
  \bibinfo{author}{\bibfnamefont{L.}~\bibnamefont{Henrard}},
  \bibinfo{journal}{Phys. Rev. Lett.} \textbf{\bibinfo{volume}{97}},
  \bibinfo{pages}{036803} (\bibinfo{year}{2006}).

\bibitem[{\citenamefont{Partoens and Peeters}(2006)}]{partoens2006graphene}
\bibinfo{author}{\bibfnamefont{B.}~\bibnamefont{Partoens}} \bibnamefont{and}
  \bibinfo{author}{\bibfnamefont{F.}~\bibnamefont{Peeters}},
  \bibinfo{journal}{Physical Review B} \textbf{\bibinfo{volume}{74}},
  \bibinfo{pages}{075404} (\bibinfo{year}{2006}).

\bibitem[{\citenamefont{Lu et~al.}(2006{\natexlab{a}})\citenamefont{Lu, Chang,
  Huang, Chen, and Lin}}]{lu2006influence}
\bibinfo{author}{\bibfnamefont{C.}~\bibnamefont{Lu}},
  \bibinfo{author}{\bibfnamefont{C.}~\bibnamefont{Chang}},
  \bibinfo{author}{\bibfnamefont{Y.}~\bibnamefont{Huang}},
  \bibinfo{author}{\bibfnamefont{R.}~\bibnamefont{Chen}}, \bibnamefont{and}
  \bibinfo{author}{\bibfnamefont{M.}~\bibnamefont{Lin}},
  \bibinfo{journal}{Physical Review B} \textbf{\bibinfo{volume}{73}},
  \bibinfo{pages}{144427} (\bibinfo{year}{2006}{\natexlab{a}}).

\bibitem[{\citenamefont{Aoki and Amawashi}(2007)}]{aoki-amawashi-2007}
\bibinfo{author}{\bibfnamefont{M.}~\bibnamefont{Aoki}} \bibnamefont{and}
  \bibinfo{author}{\bibfnamefont{H.}~\bibnamefont{Amawashi}},
  \bibinfo{journal}{Solid State Commun.} \textbf{\bibinfo{volume}{142}},
  \bibinfo{pages}{123} (\bibinfo{year}{2007}).

\bibitem[{\citenamefont{Koshino and Ando}(2007)}]{koshino-ando07}
\bibinfo{author}{\bibfnamefont{M.}~\bibnamefont{Koshino}} \bibnamefont{and}
  \bibinfo{author}{\bibfnamefont{T.}~\bibnamefont{Ando}},
  \bibinfo{journal}{Phys. Rev. B} \textbf{\bibinfo{volume}{76}},
  \bibinfo{pages}{085425} (\bibinfo{year}{2007}).

\bibitem[{\citenamefont{Koshino and Ando}(2008)}]{koshino-ando08}
\bibinfo{author}{\bibfnamefont{M.}~\bibnamefont{Koshino}} \bibnamefont{and}
  \bibinfo{author}{\bibfnamefont{T.}~\bibnamefont{Ando}},
  \bibinfo{journal}{Phys. Rev. B} \textbf{\bibinfo{volume}{77}},
  \bibinfo{pages}{115313} (\bibinfo{year}{2008}).

\bibitem[{\citenamefont{Koshino and
  McCann}(2009{\natexlab{a}})}]{koshino-aba-2009}
\bibinfo{author}{\bibfnamefont{M.}~\bibnamefont{Koshino}} \bibnamefont{and}
  \bibinfo{author}{\bibfnamefont{E.}~\bibnamefont{McCann}},
  \bibinfo{journal}{Phys. Rev. B} \textbf{\bibinfo{volume}{79}},
  \bibinfo{pages}{125443} (\bibinfo{year}{2009}{\natexlab{a}}).

\bibitem[{\citenamefont{Koshino and Ando}(2009)}]{koshino2009electronic}
\bibinfo{author}{\bibfnamefont{M.}~\bibnamefont{Koshino}} \bibnamefont{and}
  \bibinfo{author}{\bibfnamefont{T.}~\bibnamefont{Ando}},
  \bibinfo{journal}{Solid State Communications} \textbf{\bibinfo{volume}{149}},
  \bibinfo{pages}{1123} (\bibinfo{year}{2009}).

\bibitem[{\citenamefont{Koshino and McCann}(2011)}]{koshino-LL-2011}
\bibinfo{author}{\bibfnamefont{M.}~\bibnamefont{Koshino}} \bibnamefont{and}
  \bibinfo{author}{\bibfnamefont{E.}~\bibnamefont{McCann}},
  \bibinfo{journal}{Phys. Rev. B} \textbf{\bibinfo{volume}{83}},
  \bibinfo{pages}{165443} (\bibinfo{year}{2011}).

\bibitem[{\citenamefont{Sena et~al.}(2011)\citenamefont{Sena, Pereira, Peeters,
  and Farias}}]{sena-trilayer}
\bibinfo{author}{\bibfnamefont{S.~H.~R.} \bibnamefont{Sena}},
  \bibinfo{author}{\bibfnamefont{J.~M.} \bibnamefont{Pereira}},
  \bibinfo{author}{\bibfnamefont{F.~M.} \bibnamefont{Peeters}},
  \bibnamefont{and} \bibinfo{author}{\bibfnamefont{G.~A.}
  \bibnamefont{Farias}}, \bibinfo{journal}{Phys. Rev. B}
  \textbf{\bibinfo{volume}{84}}, \bibinfo{pages}{205448}
  (\bibinfo{year}{2011}).

\bibitem[{\citenamefont{Taychatanapat et~al.}(2011)\citenamefont{Taychatanapat,
  Watanabe, Taniguchi, and Jarillo-Herrero}}]{jarillo-trilayer}
\bibinfo{author}{\bibfnamefont{T.}~\bibnamefont{Taychatanapat}},
  \bibinfo{author}{\bibfnamefont{K.}~\bibnamefont{Watanabe}},
  \bibinfo{author}{\bibfnamefont{T.}~\bibnamefont{Taniguchi}},
  \bibnamefont{and}
  \bibinfo{author}{\bibfnamefont{P.}~\bibnamefont{Jarillo-Herrero}},
  \bibinfo{journal}{Nature Phys.} \textbf{\bibinfo{volume}{7}},
  \bibinfo{pages}{621} (\bibinfo{year}{2011}).

\bibitem[{\citenamefont{Lu et~al.}(2006{\natexlab{b}})\citenamefont{Lu, Lin,
  Hwang, Wang, Lin, and Chang}}]{lu-2006-abc}
\bibinfo{author}{\bibfnamefont{C.~L.} \bibnamefont{Lu}},
  \bibinfo{author}{\bibfnamefont{H.~C.} \bibnamefont{Lin}},
  \bibinfo{author}{\bibfnamefont{C.~C.} \bibnamefont{Hwang}},
  \bibinfo{author}{\bibfnamefont{J.}~\bibnamefont{Wang}},
  \bibinfo{author}{\bibfnamefont{M.~F.} \bibnamefont{Lin}}, \bibnamefont{and}
  \bibinfo{author}{\bibfnamefont{C.~P.} \bibnamefont{Chang}},
  \bibinfo{journal}{Appl. Phys. Lett.} \textbf{\bibinfo{volume}{89}},
  \bibinfo{pages}{221910} (\bibinfo{year}{2006}{\natexlab{b}}).

\bibitem[{\citenamefont{Koshino and
  McCann}(2009{\natexlab{b}})}]{koshino-abc-2009}
\bibinfo{author}{\bibfnamefont{M.}~\bibnamefont{Koshino}} \bibnamefont{and}
  \bibinfo{author}{\bibfnamefont{E.}~\bibnamefont{McCann}},
  \bibinfo{journal}{Phys. Rev. B} \textbf{\bibinfo{volume}{80}},
  \bibinfo{pages}{165409} (\bibinfo{year}{2009}{\natexlab{b}}).

\bibitem[{\citenamefont{Zhang et~al.}(2010)\citenamefont{Zhang, Sahu, Min, and
  MacDonald}}]{fang-abc10}
\bibinfo{author}{\bibfnamefont{F.}~\bibnamefont{Zhang}},
  \bibinfo{author}{\bibfnamefont{B.}~\bibnamefont{Sahu}},
  \bibinfo{author}{\bibfnamefont{H.}~\bibnamefont{Min}}, \bibnamefont{and}
  \bibinfo{author}{\bibfnamefont{A.~H.} \bibnamefont{MacDonald}},
  \bibinfo{journal}{Phys. Rev. B} \textbf{\bibinfo{volume}{82}},
  \bibinfo{pages}{035409} (\bibinfo{year}{2010}).

\bibitem[{\citenamefont{Bao et~al.}(2011)\citenamefont{Bao, Jing, Velasco~Jr,
  Lee, Liu, Tran, Standley, Aykol, Cronin, Smirnov et~al.}}]{bao-abc2011}
\bibinfo{author}{\bibfnamefont{W.}~\bibnamefont{Bao}},
  \bibinfo{author}{\bibfnamefont{L.}~\bibnamefont{Jing}},
  \bibinfo{author}{\bibfnamefont{J.}~\bibnamefont{Velasco~Jr}},
  \bibinfo{author}{\bibfnamefont{Y.}~\bibnamefont{Lee}},
  \bibinfo{author}{\bibfnamefont{G.}~\bibnamefont{Liu}},
  \bibinfo{author}{\bibfnamefont{D.}~\bibnamefont{Tran}},
  \bibinfo{author}{\bibfnamefont{B.}~\bibnamefont{Standley}},
  \bibinfo{author}{\bibfnamefont{M.}~\bibnamefont{Aykol}},
  \bibinfo{author}{\bibfnamefont{S.}~\bibnamefont{Cronin}},
  \bibinfo{author}{\bibfnamefont{D.}~\bibnamefont{Smirnov}},
  \bibnamefont{et~al.}, \bibinfo{journal}{Nature Physics}
  \textbf{\bibinfo{volume}{7}}, \bibinfo{pages}{948} (\bibinfo{year}{2011}).

\bibitem[{\citenamefont{Charlier et~al.}(1991)\citenamefont{Charlier, Gonze,
  and Michenaud}}]{graphite-abinitio}
\bibinfo{author}{\bibfnamefont{J.}~\bibnamefont{Charlier}},
  \bibinfo{author}{\bibfnamefont{X.}~\bibnamefont{Gonze}}, \bibnamefont{and}
  \bibinfo{author}{\bibfnamefont{J.}~\bibnamefont{Michenaud}},
  \bibinfo{journal}{Phys. Rev. B} \textbf{\bibinfo{volume}{43}},
  \bibinfo{pages}{4579} (\bibinfo{year}{1991}).

\bibitem[{\citenamefont{Dresselhaus and
  Dresselhaus}(2002)}]{dresselhaus-graphite02}
\bibinfo{author}{\bibfnamefont{M.}~\bibnamefont{Dresselhaus}} \bibnamefont{and}
  \bibinfo{author}{\bibfnamefont{G.}~\bibnamefont{Dresselhaus}},
  \bibinfo{journal}{Advances in Physics} \textbf{\bibinfo{volume}{51}},
  \bibinfo{pages}{1} (\bibinfo{year}{2002}).

\bibitem[{\citenamefont{McClure}(1957)}]{mcclure-gamma4-1}
\bibinfo{author}{\bibfnamefont{J.~W.} \bibnamefont{McClure}},
  \bibinfo{journal}{Phys. Rev.} \textbf{\bibinfo{volume}{108}},
  \bibinfo{pages}{612} (\bibinfo{year}{1957}).

\bibitem[{\citenamefont{McClure}(1960)}]{mcclure-gamma4-2}
\bibinfo{author}{\bibfnamefont{J.~W.} \bibnamefont{McClure}},
  \bibinfo{journal}{Phys. Rev.} \textbf{\bibinfo{volume}{119}},
  \bibinfo{pages}{606} (\bibinfo{year}{1960}).

\bibitem[{\citenamefont{Koshino and Ando}(2006)}]{koshino2006transport}
\bibinfo{author}{\bibfnamefont{M.}~\bibnamefont{Koshino}} \bibnamefont{and}
  \bibinfo{author}{\bibfnamefont{T.}~\bibnamefont{Ando}},
  \bibinfo{journal}{Physical Review B} \textbf{\bibinfo{volume}{73}},
  \bibinfo{pages}{245403} (\bibinfo{year}{2006}).

\bibitem[{\citenamefont{Orlita et~al.}(2012)\citenamefont{Orlita, Neugebauer,
  Faugeras, Barra, Potemski, Pellegrino, and Basko}}]{orlita-graphite12}
\bibinfo{author}{\bibfnamefont{M.}~\bibnamefont{Orlita}},
  \bibinfo{author}{\bibfnamefont{P.}~\bibnamefont{Neugebauer}},
  \bibinfo{author}{\bibfnamefont{C.}~\bibnamefont{Faugeras}},
  \bibinfo{author}{\bibfnamefont{A.-L.} \bibnamefont{Barra}},
  \bibinfo{author}{\bibfnamefont{M.}~\bibnamefont{Potemski}},
  \bibinfo{author}{\bibfnamefont{F.~M.~D.} \bibnamefont{Pellegrino}},
  \bibnamefont{and} \bibinfo{author}{\bibfnamefont{D.~M.} \bibnamefont{Basko}},
  \bibinfo{journal}{Phys. Rev. Lett.} \textbf{\bibinfo{volume}{108}},
  \bibinfo{pages}{017602} (\bibinfo{year}{2012}).

\bibitem[{\citenamefont{Velasco~Jr et~al.}(2012)\citenamefont{Velasco~Jr, Jing,
  Bao, Lee, Kratz, Aji, Bockrath, Lau, Varma, Stillwell et~al.}}]{velasco2012}
\bibinfo{author}{\bibfnamefont{J.}~\bibnamefont{Velasco~Jr}},
  \bibinfo{author}{\bibfnamefont{L.}~\bibnamefont{Jing}},
  \bibinfo{author}{\bibfnamefont{W.}~\bibnamefont{Bao}},
  \bibinfo{author}{\bibfnamefont{Y.}~\bibnamefont{Lee}},
  \bibinfo{author}{\bibfnamefont{P.}~\bibnamefont{Kratz}},
  \bibinfo{author}{\bibfnamefont{V.}~\bibnamefont{Aji}},
  \bibinfo{author}{\bibfnamefont{M.}~\bibnamefont{Bockrath}},
  \bibinfo{author}{\bibfnamefont{C.}~\bibnamefont{Lau}},
  \bibinfo{author}{\bibfnamefont{C.}~\bibnamefont{Varma}},
  \bibinfo{author}{\bibfnamefont{R.}~\bibnamefont{Stillwell}},
  \bibnamefont{et~al.}, \bibinfo{journal}{Nature Nanotechnology}
  \textbf{\bibinfo{volume}{7}}, \bibinfo{pages}{156} (\bibinfo{year}{2012}).

\bibitem[{\citenamefont{Morimoto et~al.}(2010)\citenamefont{Morimoto, Avishai,
  and Aoki}}]{morimoto-ac-scaling}
\bibinfo{author}{\bibfnamefont{T.}~\bibnamefont{Morimoto}},
  \bibinfo{author}{\bibfnamefont{Y.}~\bibnamefont{Avishai}}, \bibnamefont{and}
  \bibinfo{author}{\bibfnamefont{H.}~\bibnamefont{Aoki}},
  \bibinfo{journal}{Phys. Rev. B} \textbf{\bibinfo{volume}{82}},
  \bibinfo{pages}{081404} (\bibinfo{year}{2010}).

\bibitem[{\citenamefont{Morimoto and
  Aoki}(2012)}]{morimoto-two-parameter-scaling}
\bibinfo{author}{\bibfnamefont{T.}~\bibnamefont{Morimoto}} \bibnamefont{and}
  \bibinfo{author}{\bibfnamefont{H.}~\bibnamefont{Aoki}},
  \bibinfo{journal}{Phys. Rev. B} \textbf{\bibinfo{volume}{85}},
  \bibinfo{pages}{165445} (\bibinfo{year}{2012}).

\bibitem[{\citenamefont{Kawarabayashi et~al.}(2012)\citenamefont{Kawarabayashi,
  Hatsugai, and Aoki}}]{kawarabayashi-chiral-bilayer}
\bibinfo{author}{\bibfnamefont{T.}~\bibnamefont{Kawarabayashi}},
  \bibinfo{author}{\bibfnamefont{Y.}~\bibnamefont{Hatsugai}}, \bibnamefont{and}
  \bibinfo{author}{\bibfnamefont{H.}~\bibnamefont{Aoki}},
  \bibinfo{journal}{Phys. Rev. B} \textbf{\bibinfo{volume}{85}},
  \bibinfo{pages}{165410} (\bibinfo{year}{2012}).

\end{thebibliography}

\end{document}